# Kinship can hinder cooperation in heterogeneous populations

Boyu Zhang [1,†,*], Yali Dong [2,†], Cheng-Zhong Qin [3,*], Sergey Gavrilets [4,*]

[1] Laboratory of Mathematics and Complex Systems, Ministry of Education, School of Mathematical Sciences, Beijing Normal University, Beijing 100875, People's Republic of China;

[2] School of Systems Science, Beijing Normal University, Beijing 100875, People's Republic of China;

[3] Department of Economics, University of California, Santa Barbara, CA 93106, USA;

[4] Center for the Dynamics of Social Complexity, Department of Ecology and Evolution, Department of Mathematics, University of Tennessee, Knoxville, TN 37996 USA

†: These authors contributed equally to this work.

*: To whom correspondence may be addressed. Email: zhangby@bnu.edu.cn (Boyu Zhang), qin@ucsb.edu (Cheng-Zhong Qin), gavrila@utk.edu (Sergey Gavrilets)

**Abstract**

Kin selection and direct reciprocity are two most basic mechanisms for promoting cooperation in human society. Generalizing the standard models of the multi-player Prisoner's Dilemma and the Public Goods games for heterogeneous populations, we study the effects of genetic relatedness on cooperation in the context of repeated interactions. Two sets of interrelated results are established: a set of analytical results focusing on the subgame perfect equilibrium and a set of agent-based simulation results based on an evolutionary game model. We show that in both cases increasing genetic relatedness does not always facilitate cooperation. Specifically, kinship can hinder the effectiveness of reciprocity in two ways. First, the condition for sustaining cooperation through direct reciprocity is harder to satisfy when relatedness increases in an intermediate range. Second, full cooperation is impossible to sustain for a medium-high range of relatedness values. Moreover, individuals with low cost-benefit ratios can end up with lower payoffs than their groupmates with high cost-benefit ratios. Our results point to the importance of explicitly accounting for within-population heterogeneity when studying the evolution of cooperation.

**Introduction**

Understanding conditions promoting cooperation is of fundamental importance both in economics and evolutionary biology. A number of mechanisms explaining the prevalence of cooperation in human society have been proposed and studied (Nowak, 2006b; McElreath and Boyd, 2007). These include kin (Hamilton, 1964) and group selection (Wilson, 1975; Traulsen and Nowak, 2006), direct (Trivers, 1971; Axelrod, 1984) and indirect reciprocity (Nowak and Sigmund, 2005), selective incentive and punishment (Olson, 1965; Boyd and Richerson, 1992; Fehr and Gächter, 2000, 2002; Boyd et al., 2003), social norms (Ensminger and Henrich, 2014; Gavrilets and Richerson, 2017; Gavrilets, 2020) and institutions (Ostrom, 1990; Sigmund et al., 2010; Gavrilets and Duwal Shrestha, 2021). The standard mathematical models for studying these mechanisms are the Prisoner's Dilemma (PD) game and the Public Goods Game (PGG) (Rapoport and Chammah, 1965; Axelrod and Dion, 1988; Nowak, 2006a; Sigmund, 2010).

In this paper, we focus on two most basic mechanisms for promoting cooperation: kin selection and direct reciprocity. The kin selection theory relies on the concept of inclusive fitness (Nowak, 2006b; Hamilton, 1964; Hines and Maynard Smith, 1979; Rousset, 2004; Lehmann and Keller, 2006; Van Veelen, 2009; Akçay and Van Cleve, 2012). When evaluating the inclusive fitness of a behaviour (induced by a certain gene), one should include the effect of the behaviour on individuals who might carry the same gene. According to the *Hamilton rule*, natural selection can favor cooperation if the relatedness between the donor and the recipient is greater than the cost-benefit ratio (the inverse of productivity) of the donation behaviour (Hamilton, 1964). Direct Reciprocity, in contrast to the kin selection theory, offers a mechanism by which cooperation among unrelated individuals can be promoted (Trivers, 1971). In the context of direct reciprocity, individuals engage in repeated interactions, facing the risk of losing the benefits of cooperation or even facing punishment in future interactions if they fail to cooperate at present. Previous studies have identified various strategies that can sustain cooperation under repeated interactions, including GRIM, Tit-for-Tat, Win-Stay Lose-Shift, All-or-None, and the generous zero-determinant strategies (Axelrod, 1984; Sigmund, 2010; Nowak and Sigmund, 1992, 1993; Press and Dyson, 2012; Hilbe et al., 2014; Pinheiro et al., 2014; Hilbe et al., 2017). While these strategies differ significantly, a fundamental rule applies to all: direct reciprocity can foster cooperation only when the likelihood of one more round of interaction between the same individuals surpasses the cost-benefit ratio (Nowak, 2006b).

In social insects and other group-living animals, individuals within a group are not only genetically related but also have long-term relationships. This is common in humans too, e.g., in clan societies or family businesses (Bernheim and Stark, 1988). While both kin selection and direct reciprocity are vital in promoting the evolution of cooperation, the interactions between these two mechanisms have received limited analysis. Several studies have suggested that genetic relatedness can amplify the impact of direct reciprocity, particularly in homogeneous populations (Lehmann and Keller, 2006; McGlothlin et al., 2010; Van Veelen et al., 2012; Van Cleve and Akçay, 2014). However, the dynamics of heterogenous populations in relation to these factors remain unclear. This paper aims to fill this gap by presenting a model that explores repeated social interactions between genetically related individuals in the Prisoner's Dilemma (PD) game and Public Goods Game (PGG). An important and innovative aspect of our analysis is that we explicitly consider the variations among individuals in parameters that influence their costs and benefits. These parameters encompass factors like internal state, physical strength, previous experience, and more. It is noteworthy that such variations are widespread in both natural and human populations, yet they are often overlooked in existing game-theoretic models (see Gavrilets, 2015 for a recent review). As shown both empirically and theoretically (Gavrilets, 2015; Cherry et al., 2005; Tavoni et al., 2011; Nishi et al., 2015; Gavrilets and Fortunato, 2014; Hauser et al., 2019), within-group heterogeneity can be crucial for the success of cooperation.

We note that strategies in repeated PD games and PGG can be very complex. In general, there is a great multiplicity of Nash equilibria (NE in short) that can correspond to different cooperation rates, but none of which is evolutionarily stable (Hauser et al., 2019; Fudenberg and Tirole, 1991). Thus, the predictive power of Nash equilibrium or ESS in such settings is weak, especially for a heterogeneous population. A recent study showed that the evolutionary outcome in the asymmetric repeated PGG can be described by a subgame perfect equilibrium (SPE), where the group contributions are higher when full cooperation is the outcome of a SPE (Hauser et al., 2019). As a refinement of Nash equilibrium, a SPE requires that players' strategies when restricted to any subgame form a Nash equilibrium for the subgame. SPE often arise in the context of fully rational players who can perform backward induction and understand the consequences of their actions in future interactions. It can also be relevant in situations where players are not fully rational but are influenced by lower-level biological processes such as learning, imitation, or evolutionary dynamics via selection and mutation. Indeed, SPE is a useful tool for understanding how different strategies might fare in the long run, given the interactions and choices made by other individuals in evolutionary biology. In addition, SPE-based methods have also been used in the context of cooperation (Hilbe et al., 2017; Hauser et al., 2019), territorial contest (Maynard Smith, 1982), parental investment (Cressman, 2003; Broom and Rychtár, 2013), and the interaction between brood parasite and its host (Planque et al., 2002; Harrison and Broom, 2009).

In this paper, we employ the method of subgame perfection to examine the evolution of cooperation, and we validate our approach through evolutionary simulations. We establish two sets of results. The first is a set of analytical results focusing on the SPE of the asymmetric repeated PD game and PGG. In order to substantiate the validity of our approach and gain further insights into evolutionary dynamics, we conduct agent-based simulations of an evolutionary game model (Ohtsuki, 2010). Both the subgame perfect equilibrium analysis and the simulation results indicate that increasing genetic relatedness does not consistently promote cooperation in repeated interactions with diverse individuals. Additionally, it is evident that sustaining full cooperation is not viable across a range of relatedness values.

To facilitate our analytical derivations, we make the assumption that players exhibit perfect rationality. We do not impose any constraints on the strategy set and concentrate on strategies that can sustain full cooperation. Our proof establishes that the GRIM strategy plays a crucial role in sustaining cooperation. Specifically, if cooperation is achieved by all players in every round as a Subgame Perfect Equilibrium (SPE) outcome, it is only possible if the same outcome is achieved when all players adopt the GRIM strategy as their SPE. Therefore, to ensure the sustainability of full cooperation as an SPE outcome, it is sufficient to have the GRIM strategy as a SPE strategy for all players. In the context of homogeneous populations, the promotion of cooperation is consistently facilitated by increasing genetic relatedness. This effect arises from the relaxation of conditions required to sustain cooperation. In comparison, with heterogeneous populations where individuals exhibit varying cost-benefit ratios, increasing genetic relatedness within an intermediate range strengthens the conditions necessary to sustain cooperation. Consequently, it hinders cooperation instead of facilitating it. The reason is that cooperation can increase inclusive fitness for individuals with low cost-benefit ratios (or higher productivities) even in the one-shot interaction. As a result, unconditional cooperation (AllC) becomes a dominant strategy for them, which weakens the punishment facing defectors. Therefore, individuals with high cost-benefit ratios (or lower productivities) will adopt defection and full cooperation cannot be sustained.

We test whether the above surprising results can be observed in a standard evolutionary model. We show that cooperation can be sustained in the 2-person asymmetric repeated PD game without any restriction on the strategies if it can be sustained in the corresponding evolutionary game model with strategies restricted to AllC, AllD, and GRIM. It follows that for our evolutionary agent-based simulations, we can focus on 2-person games with a limited set of strategies. Following Ohtsuki (Ohtsuki, 2010), we study the classical Wright's island model (Wright, 1931) with heterogeneous individuals and the above three strategies. Our simulation results show that the SPE analysis can correctly predict the evolutionary outcome. Specifically, for high dispersal rates (which imply low relatedness), GRIM is the most frequently used strategy for all types of individuals. For intermediate dispersal rates (which imply intermediate relatedness), most individuals with low cost-benefit ratio adopt AllC and most individuals with high cost-benefit ratio adopt AllD. Moreover, these results are robust to changes in cost-benefit ratios, initial strategy distribution, the number of islands, the size of each island, and small mutation rates.

**Analytical results**

In this section, we begin by analyzing a class of 2-person asymmetric infinitely repeated PD games in the presence of relatedness. We then extend the analysis to the *n*-person case. Finally, we conduct similar analysis for a class of heterogeneous PGG. In this paper, we apply the method of subgame perfection to the evolution of cooperation. Our focus is on the sustainability of cooperation with the presence of genetic relatedness or altruism. We show using a direct extension of the method in (Hauser et al., 2019) that for our *n*-person PD game and PGG without kinship or with low enough kinship, cooperation in each round by all players is the outcome of a SPE if and only if all players play the GRIM strategy is a SPE (see Supplementary Materials, SM, Appendix A, Propositions 1 and 3). Furthermore, for games with intermediate or high kinship, assuming each player plays AllC whenever C the single-round dominant strategy, cooperation by all players in each round is achievable in SPE if and only if it is achievable in the SPE with all other players whose single-round dominant strategy is D play the GRIM strategy (see SM, Appendix A, Propositions 2 and 4). For these reasons, we only need to focus on the conditions under which the GRIM strategy is a SPE strategy in later discussions.

**The 2-person asymmetric PD game.** We first consider a simple 2-person asymmetric PD game, where C is the action to confer a benefit to the co-player at a cost and D is the action not to confer any benefit (Nowak, 2006a, 2006b). This game is also known as the 'Donation game' (Wilson, 1975; Sigmund, 2010). The benefit and cost are measured in units of payoffs. Let $b_i$ ($b_i > 0$) denote the benefit player *i* can confer to the co-player and $c_i$ ($c_i > 0$) the cost incurred by player *i* in conferring the benefit. Let $\gamma_i$ ($\gamma_i = c_i/b_i$) denote the cost-benefit ratio, whose inverse measures the productivity of player *i*. Without loss of generality, we assume $0 < \gamma_1 < \gamma_2 < 1$. That is, player 1 has a lower cost-benefit ratio (or a higher productivity) than player 2.

Suppose that the relatedness between the two players is *r*, which is the probability of sharing a gene. There are several ways to define the inclusive fitness. A traditional way is to calculate the marginal fitness effects of the focal player on itself and on other players weighted by their relatedness to the focal player (3). In the 2-person asymmetric PD game, the traditional inclusive fitness for player *i* can be given by $-c_i + r\, b_i$, and strategy C (resp. strategy D) is strictly dominant if and only if $r > \gamma_i$ (resp. $r < \gamma_i$) (Hamilton, 1964; Rousset, 2004). Thus, if the game is played only once, in NE both players defect if $r < \gamma_1$ (low relatedness), player 1 cooperates and player 2 defects if $\gamma_1 < r < \gamma_2$ (intermediate relatedness) and both players cooperate if $r > \gamma_2$, (high relatedness).

We assume that at the end of each round, the PD game is repeated for another round with probability $\omega > 0$. Note that the expected number of rounds is $1/(1-\omega)$ and the probability to stop at the end of round *t* is $(1-\omega)\omega^{t-1}$ for *t* =

1, 2, ... (Sigmund, 2010; Luce and Raiffa, 1957). This repeated game with a probabilistic end is equivalent to the one-shot game being infinitely repeated with $\omega$ as the discount factor.

Our analysis will be based on an alternative notion of inclusive fitness which we define as a sum of two components: the (direct) payoff to the focal individual and the payoff to the genetically related partner multiplied by the relatedness (e.g., Hines and Maynard Smith, 1979)[1]. Although this approach can under some conditions lead to erroneous results (Grafen, 1979), it allows for a significant analytical advance in the case of repeated interactions without any restriction on the strategy set.

For the 2-person one-shot PD game, our inclusive payoff matrix is shown in Table 1. In this game, strategy C (resp. strategy D) is strictly dominant for player $i$ if and only if $-c_i + rb_i > 0$ (or equivalently, $r > \gamma_i$). Thus, the outcome is the same as the above analysis based on the traditional inclusive fitness. In addition, we show in SM Appendix B that for the 2-person repeated PD game, the results of the SPE analysis based on alternative definition without any restriction on possible strategies are the same as the results of the invasion analysis based on traditional definition with three limited strategies, AllC, AllD, and GRIM. This in turn provides evidence that our SPE analysis is able to identify general patterns and results.

**Table 1. Inclusive payoff matrix for the one-shot 2-person PD game with relatedness $r$.** The first and the second entries are payoffs for player 1 and player 2, respectively. The terms in blue are players' own payoff and those in red are the payoffs obtained from relatives through relatedness.

|  |  | Player 2 | |
|---|---|---|---|
|  |  | C | D |
| Player 1 | C | $b_2 - c_1 + r(b_1 - c_2), b_1 - c_2 + r(b_2 - c_1)$ | $-c_1 + rb_1, b_1 - rc_1$ |
|  | D | $b_2 - rc_2, -c_2 + rb_2$ | 0,0 |

Let $u_t$ be the inclusive payoff received in round $t$, then the expected payoff is $(1-\omega)\sum_{t=1}^{\infty}\omega^{t-1}u_t$. A player using the GRIM strategy chooses C in the first round and continues to choose C as long as both players chose C in all previous rounds; if a player deviated in a previous round, then switch to D in the next and each of future rounds. Refusing to cooperate is understood as punishment that would follow after defection has taken place.

In SM Appendix A, we analyze SPE for the above repeated PD game with kinship. Our results can be summarized as follows.

- Without relatedness $r = 0$, choosing D in each round (AllD) is a SPE since D is the one-shot strictly dominant strategy for both players. Furthermore, GRIM is a SPE for player $i$ when $\omega > \frac{c_i}{b_j}$. Thus, the critical value of $\omega$ for sustaining cooperation via the GRIM is $\omega^*(0) = max\left\{\frac{c_2}{b_1}, \frac{c_1}{b_2}\right\}$ (to be referred to as the critical probability). In the homogenous case (i.e., when $b_1 = b_2 = b$, $c_1 = c_2 = c$), the condition for sustaining cooperation takes the form of

---

[1] It is worth remarking that our inclusive fitness function can be viewed as a linear form of altruistic utility function from the perspective of behavioral economics, where the coefficient of relatedness can be interpreted as representing a degree of altruism. With this alternative interpretation, our models involve genetically unrelated individuals who are altruistic towards each another (see detailed explanations in Discussion).

- $\omega \geq c/b$. In words, the probability to play another round should be sufficiently large (1). In particular, this result is the same as in the case of traditional inclusive fitness where players have two possible strategies GRIM and AllD.
- With low relatedness $r < \gamma_1$, the one-shot strictly dominant strategy is still D for both players, but in this case, the critical probability $\omega$ which can sustain mutual cooperation in repeated play via the GRIM is $\omega^*(r) = max\left\{\frac{c_1 - rb_1}{b_2 - rc_2}, \frac{c_2 - rb_2}{b_1 - rc_1}\right\}$. Notice that $\frac{c_i - rb_i}{b_j - rc_j}$ decreases with $r$. It follows that $\omega^*(r)$ decreases with $r$. The smaller the critical probability is, the less restrictive the condition for sustaining cooperation through GRIM becomes. Therefore, increasing relatedness facilitates the cooperation for $r < \gamma_1$. This outcome is intuitive. In particular, this result is the same as in the case of traditional inclusive fitness where players have two possible strategies GRIM and AllD.
- With intermediate relatedness $\gamma_1 < r < \gamma_2$, the one-shot strictly dominant strategy is C for player 1 and D for player 2. In this case, GRIM cannot be a SPE strategy because player 1 has no incentive to choose D. With such intermediate relatedness, the most cooperative SPE is for player 1 to play AllC and player 2 to play AllD. As a result, mutual (full) cooperation cannot be sustained. In particular, this result is the same as in the case of traditional inclusive fitness where player 1 has AllC and AllD and player 2 has GRIM and AllD as their two available strategies, respectively.
- Finally, with high relatedness $r > \gamma_2$, the one-shot strictly dominant strategy is C for both players. Consequently, AllC is a SPE strategy for both players which sustains cooperation for any $\omega$ (i.e., the critical value of $\omega$ is 0). In particular, this result is the same as in the case of traditional inclusive fitness where players have two possible strategies AllC and AllD.

Note that in the homogeneous case with $c_1 = c_2 = c$, $b_1 = b_2 = b$ and $r_1 = r_2 = r$, the critical probability for sustaining cooperation is $\omega^*(r) = (c - rb)/(b - rc)$. This value monotonically decreases with $r \in [0, c/b]$, so that increasing relatedness always facilitates cooperation. However, if players are heterogeneous, our results imply that increasing relatedness enhances cooperation only for small ($r < \gamma_1$) or large ($r > \gamma_2$) values of relatedness $r$, and hinders cooperation for its intermediate values ($\gamma_1 < r < \gamma_2$). These results are illustrated in Figure 1.

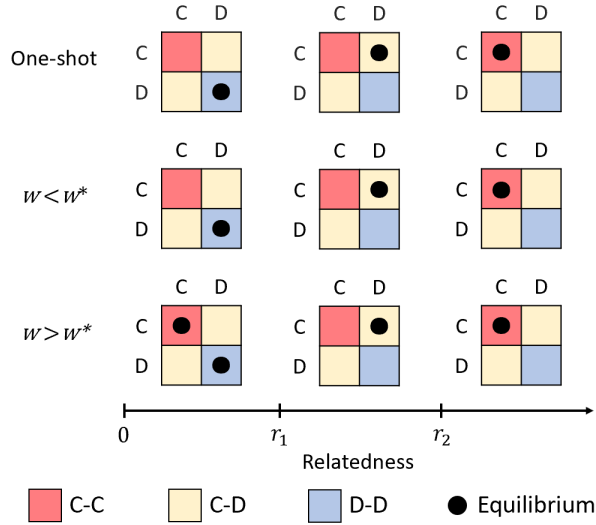

**Figure 1. Equilibrium in the 2-person repeated PD game with relatedness $r$.** For low relatedness, mutual cooperation

can be sustained in SPE via GRIM when $\omega > \omega^*(r)$ or $\omega > \omega^*(0)$. For intermediate relatedness, the most cooperative SPE is one in which one player chooses AllD while the other chooses AllC. For high relatedness, AllC is the only SPE.

**The *n*-person asymmetric PD game.** We now extend the analysis to *n*-person case. The generalization of the 2-person PD to allow for more than two players was independently established in 1973 by (Dawes, 1973; Hamburger, 1973; Schelling, 1973) (see also Hamilton, 1975; Fletcher and Zwick, 2004). This game is closely related to an "other's only" variant of the PGG, in which players do not obtain any return from their own contribution and hence, are faced with an even more pronounced social dilemma than the standard PGG (Axelrod and Dion, 1988; Sigmund, 2010; Hauser et al., 2019; Sasaki et al., 2012). The game is also related to the Mutual Aid Game in (Sugden, 1986; Zhang et al., 2014).

In an *n*-person asymmetric PD game, player $i$ can either take action C or action D. Action C confers a benefit of $b_i$ at a cost of $c_i$, where the benefit is equally distributed among the other $n-1$ players. For each player $i$, let $x_i = 0$ denote player $i$'s action D and $x_i = 1$ denote his/her action C. Given $\boldsymbol{X} = (x_1, \dots, x_n)$ with $x_i = \{0,1\}$, player $i$'s net payoff[2] obtained from the PD game is given by

$$f_i(\boldsymbol{X}) = -c_i x_i + \sum_{j \neq i} \frac{b_j x_j}{n-1}, \tag{1}$$

It is clear that $x_i = 0$ is strictly dominant for player $i$ if the game is one-shot.

With kinship $r$, player $i$'s payoff function becomes

$$f_i(\boldsymbol{X}, r) = f_i(\boldsymbol{X}) + r \sum_{j \neq i} f_j(\boldsymbol{X}) = (1-r) f_i(\boldsymbol{X}) + r \sum_{j=1}^{n} f_j(\boldsymbol{X}). \tag{2}$$

Notice that

$$\sum_{j=1}^{n} f_j(\boldsymbol{X}) = \sum_{j=1}^{n} (b_j - c_j) x_j. \tag{3}$$

It follows that

$$f_i(\boldsymbol{X}, r) = (rb_i - c_i) x_i + \sum_{j \neq i} \frac{b_j x_j}{n-1} + r \sum_{j \neq i} \left( \frac{n-2}{n-2} b_j - c_j \right) x_j. \tag{4}$$

Thus, in the one-shot PD with kinship, $x_i = 0$ is strictly dominant for player $i$ when $r < c_i/b_i$ and $x_i = 1$ is strictly dominant when $r > c_i/b_i$.

As for the 2-person case, we denote the cost-benefit ratio for individual $i$ by $\gamma_i = c_i/b_i$ and without loss of generality assume $0 < \gamma_1 < \cdots < \gamma_n < n$. Thus, without kinship, D is strictly dominant for all players if the game is played only once. With kinship, D is still strictly dominant for player $i$ when $r < \gamma_i$ and C is strictly dominant for player $i$ when $r > \gamma_i$.

As before, we assume that at the end of each round, the above *n*-person PD game is repeated for another round with probability $\omega > 0$. Similar to the analysis of the 2-person case, kinship enhances the effect of direct reciprocity for both low and high relatedness, in that kinship relaxes the critical probability (discount factor) for GRIM to be a SPE strategy when $r < \gamma_1$ and AllC becomes the strictly dominant strategy for all players when $r > \gamma_n$ (see Figure 2 and SM Appendix A). The case of intermediate relatedness is more subtle and complicated. For $\gamma_m < r < \gamma_{m+1}$ with $m \leq n-1$, the single-round dominant strategy is C for players 1 to $m$ and D for players $m+1$ to $n$. Note that the larger $r$ is, the more players will adopt AllC. However, increasing the number of AllC players also weakens the punishment, which would follow after defection has taken place. As shown in Figure 2, the critical probability $\omega^*(r)$ does not change monotonically with $r$. On the one hand, in each interval $(\gamma_m, \gamma_{m+1})$, $\omega^*(r)$ decreases with $r$ for $1 < m \leq n-2$. On the other hand, $\omega^*(\gamma_m^+) > \omega^*(\gamma_m^-)$ for $1 < m \leq n-1$, i.e., $\omega^*(r)$ jumps up as $r$ increases from $\gamma_i^-$ to $\gamma_i^+$. A (counterintuitive) implication is that the more individuals with cost-benefit ratios smaller than the relatedness, the harder to sustain cooperation through GRIM, in the sense that the condition on the critical probability is harder to satisfy (see detailed calculations in SM Appendix A). In particular, when $m = n-1$, D is the single-round strictly dominant strategy for player $n$ while C is the single-round dominant strategy for all other players. Consequently, to sustain cooperation in SPE, players 1 to $n-1$ will adopt AllC regardless of player $n$'s strategy. As a result, player $n$ will choose AllD and full cooperation cannot be sustained in SPE for $\gamma_{n-1} < r < \gamma_n$ (see Figure 3). This shows that relatedness can hinder cooperation.

---

[2] Players' payoffs in the 2-person case are net payoffs (incremental changes to total payoffs). For players' total payoffs, we need to include their baseline payoffs. Since baseline payoffs are independent of players' strategies in the game, they are behaviourally irrelevant. For this reason, we continue to work with players' net payoffs only for the *n*-person generalization of PD and the subsequent public goods game.

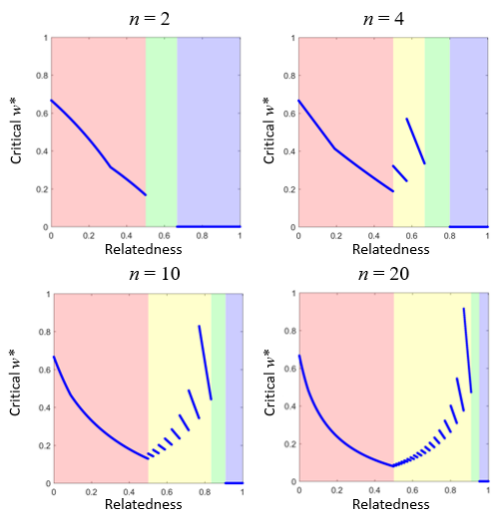

**Figure 2. The critical probability $\omega^*(r)$ in the $n$-person repeated PD game with relatedness $r$.** The blue curves denote the relationship between $\omega^*(r)$ and the relatedness that can sustain full cooperation in SPE via GRIM. Parameters are taken as $c_i = 1$ and $b_i = 2 - (i-1)/n$ for player $i = 1, 2, \cdots, n$. For low relatedness (red region), $\omega^*(r)$ is decreasing in $r$. For intermediate relatedness (yellow and green regions), a higher $r$ does not necessarily lead to a lower $\omega^*(r)$. In particular, for $\gamma_{n-1} < r < \gamma_n$ (green region), full cooperation cannot be a SPE outcome for any $\omega$. Finally, for high relatedness, AllC strategy is a SPE for all $\omega$, implying that $\omega^*(r) = 0$.

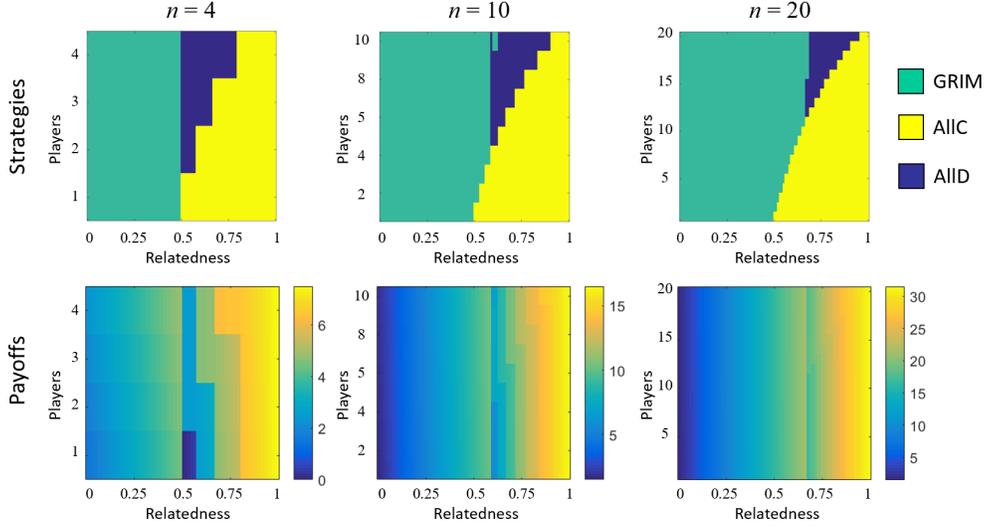

**Figure 3. Players' strategies and payoffs in the most cooperative SPE of the *n*-person repeated PD game with relatedness *r*.** Parameters: $\omega = 0.5$, $c_i = 1$ and $b_n$ to $b_1$ are uniformly distributed in interval (1, 2] with $b_i = 2 - (i-1)/n$ for player $i = 1, 2, \cdots, n$. The *n* players are arranged along the vertical axe. Upper panel: Players' strategies in SPE. For low relatedness, GRIM is a SPE strategy while for high relatedness, AllC is a SPE strategy. In contrast, for intermediate relatedness, AllC is the SPE strategy for players with lower cost-benefit ratios, and AllD is the SPE strategy for players who have higher cost-benefit ratios. Lower panel: Players' expected payoffs in SPE. For both low and high relatedness, players have similar expected payoffs. However, for intermediate relatedness, players with higher cost-benefit ratios receive greater payoffs than players with lower cost-benefit ratios because they adopt AllD.

**Heterogeneous PGG.** For a further extension of our previous analysis, we consider a heterogeneous version of the linear PGG. In this game, each group member benefits from the contributions of the group members to the common pool of a public good, so that there is an incentive to free ride. However, if all players choose to free ride, then they are all worse off (Olson, 1965). Specifically, suppose that player *i* can contribute at most $c_i$ which yields a group benefit of $b_i$. So, if player *i* contribute fraction $x_i (0 \leq x_i \leq 1)$ of $c_i$, then the group will benefit $b_i x_i$ and the total benefit of the group from the *n* members' contributions is $\sum_{i=1}^{n} b_i x_i$. Assume that the share of benefit of player *i* is $v_i$ with $\sum_{i=1}^{n} v_i = 1$, player *i*'s net payoff from the PGG is given by

$$f_i(\mathbf{X}) = (v_i b_i - c_i) x_i + v_i \sum_{j \neq i} b_j x_j, \tag{5}$$

where $\mathbf{X} = (x_1, \ldots, x_n)$ is the strategy profile. Clearly, full contribution ($x_i = 1$) is a dominant strategy for player *i* if $v_i b_i / c_i > 1$ while free-riding ($x_i = 0$) is a dominant strategy if $v_i b_i / c_i < 1$. We assume $0 \leq v_i < c_i / b_i < 1$. With these specifications, the only NE is $x_i = 0$ for all $i$. But to be socially optimal requires that each player fully contribute.

With kinship *r*, player *i*'s payoff function becomes

$$f_i(\mathbf{X}, r) = f_i(\mathbf{X}) + r \sum_{j \neq i} f_j(\mathbf{X}) = (1-r) f_i(\mathbf{X}) + r \sum_{j=1}^{n} f_j(\mathbf{X}). \tag{6}$$

Notice that

$$\sum_{j=1}^{n} f_j(\boldsymbol{X}) = \sum_{j=1}^{n}(b_j - c_j)x_j. \tag{7}$$

It follows that

$$f_i(\boldsymbol{X}, r) = (v_i b_i - c_i + r(1 - v_i)b_i)x_i + v_i \sum_{j \neq i} b_j x_j + r \sum_{j \neq i}((1 - v_j)b_j - c_j)x_j. \tag{8}$$

Similar to the PD game, kinship can promote cooperation in the one-shot PGG. For example, full contribution ($x_i = 1$) is strictly dominant for player $i$ when $r > (c_i/b_i - v_i)/(1 - v_i)$. Set $\gamma_i = (nc_i/b_i - 1)/(n - 1)$ for $i = 1, 2, \ldots, n$. Without loss of generality, we assume $0 < \gamma_1 < \cdots < \gamma_n < 1$. Thus, $x_i = 0$ is strictly dominant for player $i$ when $r < \gamma_i$ and $x_i = 1$ is strictly dominant when $r > \gamma_i$.

We further assume that the game continues for another round with probability $\omega$ at the end of each round. In this setting, an AllC player unconditionally fully contributes while an AllD player never contributes. In comparison, a GRIM player will fully contribute in the first round and continue to fully contribute as long as all players fully contributed in all previous rounds; if someone stopped fully contributing in a previous round, stop contributing in the next and each of the future rounds (Hauser et al., 2019).

We show that when the total benefit of the public good is equally shared among the $n$ players ($v_i = \frac{1}{n}$), kinship lowers the critical probability $\omega^*(r)$ when $r < \gamma_1$, and AllC becomes the strictly dominant strategy for all players when $r > \gamma_n$ (see Figure 4 and SM Appendix C). Furthermore, for the intermediate range, $\gamma_1 < r < \gamma_{n-1}$, $\omega^*(r)$ decreases with $r$ over $(\gamma_m, \gamma_{m+1})$ for $1 < m \leq n - 2$ and $\omega^*(\gamma_m^+) > \omega^*(\gamma_m^-)$ for $1 < m \leq n - 1$. Finally, full contribution cannot be sustained when $\gamma_{n-1} < r < \gamma_n$. In summary, player $i$'s SPE strategy is AllC when $\gamma_i < r$ and GRIM or AllD when $\gamma_i > r$. These results are illustrated in Figure 5. We note that SPE is more complicated when the share of benefit is heterogeneous, but our main result remains qualitatively similar: a larger $r$ does not always facilitate the evolution of cooperation (see SM Appendix D and Figure S1).

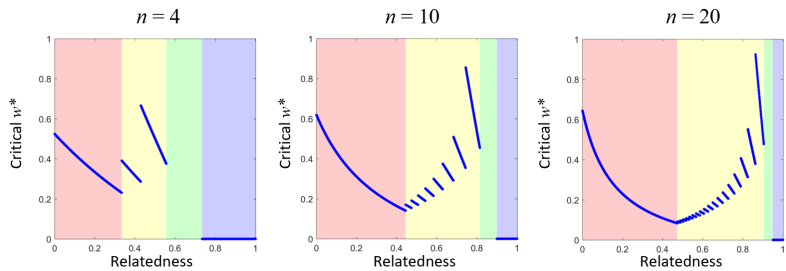

**Figure 4. The critical probability $\boldsymbol{\omega^*(r)}$ in the $\boldsymbol{n}$-person repeated heterogeneous PGG with a probabilistic end and relatedness $\boldsymbol{r}$.** The blue curves describe the relationship between $\omega^*(r)$ and r for sustaining cooperation via GRIM. Parameters are taken as $c_i = 1$ and $v_i = 1/n$ for all players, and $b_n$ to $b_1$ are uniformly distributed in interval (1, 2] with $b_i = 2 - (i - 1)/n$ for player i. For low relatedness (red region), $\omega^*(r)$ decreases with r. For intermediate relatedness (yellow and green regions), a higher $r$ does not necessarily lead to a lower $\omega^*(r)$. In particular, for $\gamma_{n-1} < r < \gamma_n$ (green region), full cooperation cannot be sustainable in SPE for any $\omega$. Finally, for high relatedness, AllC is a SPE strategy for all players and for all $\omega$ as $\omega^*(r) = 0$.

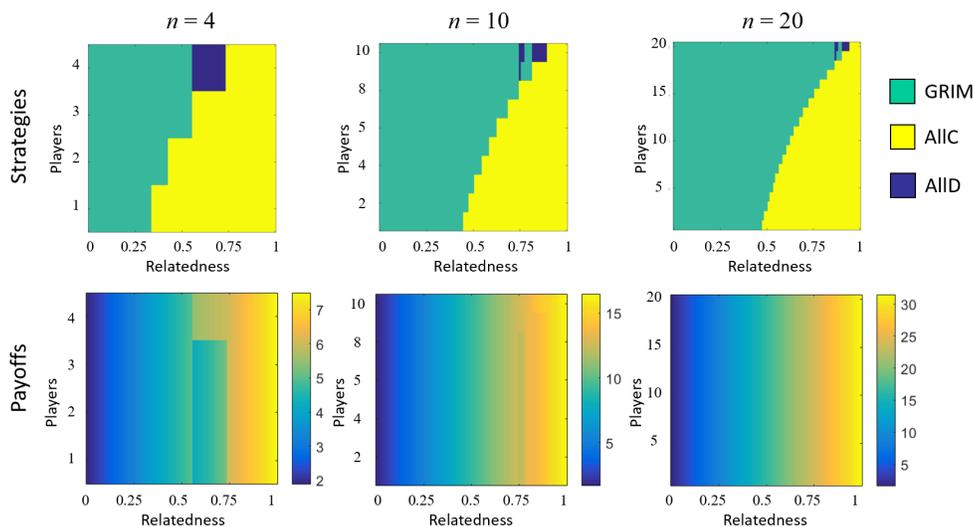

**Figure 5. Players' strategies and payoffs in the most cooperative SPE of the *n*-person repeated heterogenous PGG with a probabilistic end and relatedness *r*.** Parameters: $\omega = 0.5$ (i.e., the expected number of rounds is 2), $c_i = 1$ and $v_i = 1/n$ for all players, and $b_n$ to $b_1$ are uniformly distributed in interval (1, 2] with $b_i = 2 - (i-1)/n$ for player i. The vertical axe represents the n players. Upper panel: Players' strategies in SPE. For low relatedness, GRIM is a SPE strategy and for high relatedness, AllC is a SPE strategy. In contrast, for intermediate relatedness, AllC is the SPE strategy for players with lower cost-benefit ratios and AllD is the SPE strategy for those with high cost-benefit ratios. Lower panel: Players' expected payoffs in SPE.

**Agent-based simulations based on Wright's island model**

Our analysis above assumes that individuals are perfectly rational and focuses on the most cooperative SPE. However, the repeated PD game and PGG have infinitely many NEs and SPE. In particular, AllD is also a SPE for games with no or low relatedness. Therefore, it is not clear whether cooperation can evolve and whether the cooperation rate changes monotonically with genetic relatedness. In this section, we numerically test the robustness of our analytical results using an evolutionary game variant of Wright's island model (Ohtsuki, 2010). It is worth noting that our goal is not to exactly reproduce our analytical model and SPE results in the island model, but rather to show by agent-based simulations that our surprising results can be also observed in a standard evolutionary model.

We consider the 2-person asymmetric PD game. Suppose that there are $G$ islands, each of size $N$ haploid asexual individuals. In each generation, individuals in an island randomly form pairs and play the repeated PD game. Within each pair, one individual is randomly chosen to be player 1 (called rank 1), and the other is player 2 (called rank 2). In the SPE analysis, the condition that GRIM is a SPE is the same as the condition that GRIM weakly dominates AllD and/or AllC in the traditional inclusive fitness model. This implies that cooperation can be sustained in 2-person PD games without any restriction on the strategies if GRIM can survive in the corresponding evolutionary games model with traditional inclusive fitness and the above 3-strategy. We thus consider selection among 6 genetically determined types of players (three strategies for each of the two ranks). Following the standard practice, the payoffs obtained in the repeated game are converted into reproductive success in the form of $\exp(\lambda \cdot \text{payoff})$, where $\lambda$ measures the intensity of selection. In each island, offspring compete for $N$ vacant spots on an equal basis. Thus, the probability of an individual being selected as a parent of a vacant spot is proportional to $\exp(\lambda \cdot \text{payoff})$. Each offspring either remains in its natal island with probability $1-d$ or disperses to another randomly chosen island with probability $d$. Furthermore, mutation (to a randomly chosen strategy) happens with probability $\mu$.

Our setting is similar to the third example in Ohtsuki (Ohtsuki, 2010). The only difference is that the interaction in our model is asymmetric (we note that 51 considered TFT instead of GRIM, but GRIM and TFT yield the same payoff vs AllC or AllD). Overall, a smaller dispersal rate $d$ implies a higher within-group relatedness. When the number of islands $G \to \infty$, the genetic relatedness between two individuals, defined as the probability that two individuals randomly sampled from the same island are identical by descent, is $r = \frac{(1-d)^2}{N-(N-1)(1-d)^2}$.

For the case of symmetric interaction, Ohtsuki (Ohtsuki, 2010) shows that under weak selection, the Price equation can be approximated as the sum of three terms: direct payoff, benefit from relatives, and cost of competition among relatives. With unlimited dispersal (i.e., $d=1$ and $r=0$), the domain where direct reciprocity is favored by natural selection (i.e., the frequency of TFT increases) exists if and only if $\omega > c/b$. Furthermore, the effect of kin selection is stronger than in-group competition and relatedness promotes the evolution of direct reciprocity in the sense that limited dispersal (i.e., $d<1$ and $0<r<1$) widens this domain. This result matches our SPE analysis for the 2-person symmetric PD game. However, when the dispersal rate is very low, the competition among relatives weakens the effect of kin selection. In the limit of $d=0$ (i.e., $r=1$), the selection is neutral and the frequencies of the three strategies depends on the initial conditions (see Eq.5b in Ohtsuki, 2010). This implies that Wright's island model can be used to test our analytical results for low and intermediate range of relatedness.

Due to its restriction to symmetric interactions, Ohtsuki's method cannot be directly applied to the asymmetric PD game, and complete evolutionary analysis of equilibrium outcomes is difficult. Instead, we conduct numerical simulations for different parameter combinations, and we focus on the stable strategy. In general, increasing the relatedness (or decreasing the dispersal rate) does not always promote cooperation and direct reciprocity. For $\mu = 0$, our results show that the average frequency of cooperating acts in a repeated game between two randomly chosen individuals in a randomly chosen island (called cooperation rate for simplicity) displays a U-shape curve, with rapidly dropping from 0.7 to 0.2 in the interval $r \in [0,0.2]$ and slowly returning to 0.5 in the interval $r \in [0.2,1]$ (see the bold green curve in Figure 6 column 1). Moreover, our SPE analysis can qualitatively predict the strategies adopted by different types of players (see detailed discussions on stable strategy pairs in Appendix E). For $r=0$, GRIM is the most frequent strategy for both types of players, and (GRIM, GRIM) is the most frequent stable strategy pair. For $r>0$, different players adopt different strategies, with AllD being the most frequent strategy for player 2 and player 1 being more likely to adopt AllC as $r$ increases. In particular, (GRIM, GRIM) is no longer a stable strategy pair for intermediate relatedness and the most cooperative stable strategy pair is (AllC, AllD). This implies that full cooperation can be sustained in the evolutionary process for low relatedness, but the cooperation rate is at most one half for an intermediate range of relatedness. The above results are robust to small mutation rates (see Figure 6 columns 2-4) and variation in cost-benefit ratios, initial strategy distribution, the number of islands, and the size of each island (see Appendix Figure S3-S4).

We also investigate the emergence of cooperation in this Wright's island model under the condition that initially 90% of the individuals adopt AllD while the rest adopt GRIM (5%) and AllC (5%). For $\mu=0$, we find that direct reciprocity plays a non-trivial role in establishing cooperation in that about one-third of the players adopt GRIM for intermediate relatedness (see Appendix Figure S5). This result is consistent with previous studies showing that in homogeneous populations a small fraction of reciprocators can rapidly establish cooperation in a defective population (Nowak and Sigmund, 1992, 1993). for $\mu > 0$, the result is the same as those summarized in Figure 6, where the effect of relatedness on cooperation is non-monotonic. This demonstrates that our result is robust to initial strategy distribution when small mutations exists. More results on robustness test and the Matlab code can be find in https://volweb2.utk.edu/~gavrila/Relatedness/, where a broader range of parameter combinations is considered (see Appendix E for details).

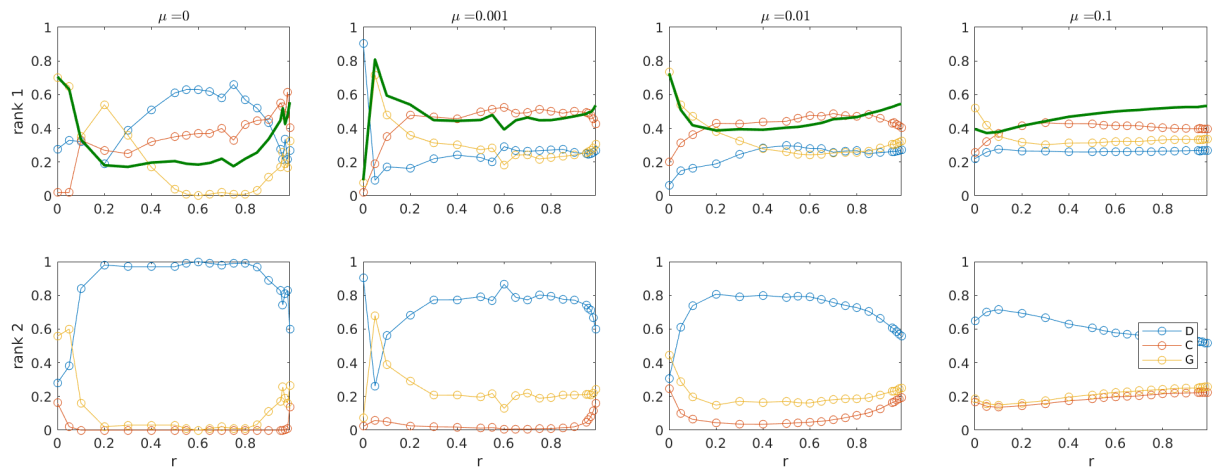

**Figure 6.** Strategy distribution in 2-person repeated PD game based on Wright's island model. These figures show the dependencies of the frequencies of strategies on $r$. The bold green lines plot the overall cooperation rates. Total number of islands is $G = 200$ with the size of each island as $N = 10$, $\omega = 0.9$, $b = 3$, $c_1 = 0.5$, $c_2 = 1$. Dispersal rate $d$ is converted into relatedness coefficient $r$ using the formula $r = \frac{(1-d)^2}{N-(N-1)(1-d)^2}$ in (Ohtsuki, 2010). Mutation rates from column 1 to column 4 are 0, 0.001, 0.01, and 0.1, respectively. For each parameter combination, we conduct 100 independent runs, where each run simulates $T = 10,000$ generations and the initial strategy distribution is random (so that the initial frequencies of AllC, AllD, and GRIM are 1/3.). We calculate the average strategy distribution over the last 100 generations over the 100 runs.

**Discussion**

It is intuitive that genetic relatedness between individuals should in general enhance cooperation. We showed that this was indeed the case with homogeneous populations, in that high relatedness relaxes the restrictions on the probability of a further encounter for sustaining mutual cooperation through GRIM. In contrast, we found that this was no longer the case when individuals are heterogeneous with respect to cost-benefit ratios. Specifically, we showed that with heterogeneity, kinship can hinder cooperation in that higher relatedness in the intermediate range makes the condition for sustaining cooperation through GRIM harder to satisfy. Furthermore, while all individuals use the same strategy with low and high relatedness (GRIM and AllC, respectively), only a mixture of AllC by some individuals and GRIM or AllD by the other individuals are sustainable in sub-game perfect equilibrium (SPE) with intermediate relatedness. Roughly, with low relatedness, direct reciprocity via GRIM plays a predominant role in sustaining cooperation. In this case, GRIM can be a SPE strategy and unilateral deviators will be credibly punished by withdrawal of cooperation. For high relatedness, on the other hand, cooperation is the one-shot dominant strategy. However, for relatedness in an intermediate range, cooperation (resp. defection) is the single round dominant strategy for individuals with low (resp. high) cost-benefit ratios. As a result, the SPE strategy at the most cooperative equilibrium is AllC for individuals with low cost-benefit ratios because punishing the partner via GRIM leads to a lower payoff than playing AllC. In contrast, the SPE strategies for individuals with high cost-benefit ratios depend on the population state: they will use AllD if many individuals use AllC (this happens if many individuals in the population have low cost-benefit ratios), and may use GRIM if only few individuals use AllC. In particular, for relatedness in the range of $\gamma_{n-1} < r < \gamma_n$, only cooperation by low cost-benefit ratio individuals can be sustainable in SPE.

We also showed that for intermediate relatedness, individuals with low cost-benefit ratios can end up with lower payoffs than their counterparts who have high cost-benefit ratios. This finding is counterintuitive within the realm of the majority of evolutionary models. What makes our approach different is that we look at games between individuals who differ not only in their strategies but also in some additional characteristics. Related results are known in both economics and evolutionary biology literature. In our models, increasing relatedness incentivize individuals with high cost-benefit ratios (or low productivity) to switch to defection because individuals with low cost-benefit ratios (or high productivity) still cooperate. This phenomenon is related to the "altruistic bully" effect in (Gavrilets, 2015; Gavrilets and Fortunato, 2014) where stronger/dominant individuals ("bullies") cooperate while the weaker or subordinate individuals defect. It is also related to Olson's (Olson, 1965) classical phenomenon of the "exploitation of the great by the small" when a public good is provided by individuals who benefits the most from it. Individuals who benefit the least from collective action can free-ride on the effort of those who potentially can benefit the most. In models of between-group competition, the "small" is exploiting the fact that the "great" is motivated to ensure the group's success. In the models considered in this paper, the "small" exploits the motivation of the "great" to increase its inclusive fitness.

Negative effects of relatedness on cooperation have been observed in experimental studies. For example, this was the case in experiments with Norway rats as found in (Schweinfurth and Taborsky, 2018). The authors speculated that this result may be due to coercion, commodity trading or correlated payoffs. Thompson et a l. (Thompson et al., 2017) observed in a study of wild banded mongooses that dominant males targeted females more closely related to them for eviction from the group. They explained this observation by selection for related subordinate individuals to submit more easily. The mechanisms investigated in this paper provides an additional possibility which is worth studying experimentally.

We note that the way we defined the inclusive fitness in the SPE analysis is different from those used in some recent work (Rousset, 2004; Akçay and Van Cleve, 2012; Taylor and Frank, 1996; Rodrigues and Gardner, 2022; Alger et al., 2020). Specifically, we used a simpler approach based on the classical work of Maynard Smith and others: the inclusive payoff of a player is the sum of his/her own payoff and the payoffs obtained from relatives (Hines and Maynard Smith, 1979; Bergstrom, 1995). Although our SPE analysis did not use the more refined definition of kinship, we show in SM Appendix B that for the 2-person PD game, the results are the same as in the case of traditional inclusive fitness with limited strategies. This in turn provides evidence that our SPE analysis is able to identify general patterns and results.

Our analytical results assumed perfectly rational individuals and analyzed SPE, which is not a standard approach in evolutionary game. To show that our conclusions remain valid in a traditional evolutionary biology set-up, we ran agent-based simulations based on the Wright's island model, which is a standard model for studying effects of kinship. We note that the evolutionary mechanisms in the agent-based simulations and the SPE analysis are different. In the former, behaviour is genetically controlled and the relatedness is endogenous while in the latter individuals are assumed to be perfectly rational and the relatedness is exogenous. However, simulation results confirm that our SPE analysis can indeed predict the evolutionary outcome in the island model. Specifically, for intermediate dispersal rates (which imply intermediate relatedness), most individuals with low cost-benefit ratios adopt AllC and most individuals with high cost-benefit ratio adopt AllD.

It is also worth remarking that the coefficient of relatedness in our inclusive fitness function can be interpreted as representing a degree of altruism (e.g., Eq.6 in Alger and Weibull, 2010). With this alternative interpretation, the models analyzed in this paper involve individuals who are altruistic towards each another without being genetically related. In behavioural economics, a general form of altruistic utility function for player $i$ is often written as $u_i(f_1, f_2)$, where $f_j$ denotes the monetary payoff for player $j$ ($j = 1,2$) and $u_i$ is assumed to be strictly increasing in $f_j$ (Alger and Weibull, 2010; Alger et al., 2020; Andreoni and Miller, 2002). Thus, our inclusive fitness can be viewed as a linear altruistic utility function (Alger and Weibull, 2010; Andreoni and Miller, 2002). Experimental studies of altruistic preferences in human population can be found in (Andreoni and Miller, 2002; Charness and Rabin, 2002).

Overall, our results show that when individuals in asymmetric social dilemma games have low relatedness, direct reciprocity via GRIM plays a predominant role in sustaining cooperation. When relatedness is high, however, kin selection plays a predominant role in sustaining cooperation because cooperation is the one-shot dominant strategy and as a result, AllC is a SPE strategy for all individuals. More importantly and surprisingly, kinship may hinder the effectiveness of reciprocity in two ways. First, the condition for sustaining cooperation through direct reciprocity is harder to satisfy as relatedness increases within an intermediate range. Second, full cooperation is impossible to sustain for a medium-high range of relatedness values. In this case, other mechanisms may be required to facilitate the evolution of cooperation. Our results point to the importance of explicitly accounting for within-population heterogeneity when studying the evolution of cooperation with game-theoretic models.

**Acknowledgements:** We thank Christian Hilbe and Karl Sigmund for helpful discussions. **Funding:** B.Z. was supported by the National Natural Science Foundation of China under Grants 72131003, 71922004 and the Beijing

Natural Science Foundation under Grant Z220001. Y.D. was supported by the National Natural Science Foundation of China under Grants 72103021, 72091511, S.G. was supported by the U. S. Army Research Office grants W911NF-14-1-0637 and W911NF-18-1-0138 and the Office of Naval Research grant W911NF-17- 1-0150, the National Institute for Mathematical and Biological Synthesis through NSF Award #EF-0830858, and the Air Force Office of Scientific Research grant FA9550-21-1-0217. **Author contributions:** B.Z., Y.D., C.Q. performed the analysis, B.Z., Y.D., C.Q., S.G. discussed the results and wrote the manuscript. **Competing interest:** The authors declare no competing interests.

**Data and materials availability:** All data needed to evaluate the conclusions in the paper are present in the paper and/or the Supplementary Materials. Data files and computing scripts may be requested from the authors.